\documentclass[aps,prl,twocolumn,amsmath,amssymb]{revtex4}

\usepackage{graphicx}
\usepackage{bm}

\begin{document}

\title{Signatures of the Anderson localization excited by an optical frequency comb}

\author{ S.Gentilini$^{1,2}$, A.Fratalocchi$^{2}$, C.Conti$^{1,3}$}

\affiliation{
$^1$Research Center Soft INFM-CNR, c/o Universit\`a di
Roma ``Sapienza,'' I-00185, Roma, Italy \\
$^2$Dipartimento di Fisica, Universit\`a di Roma ``Sapienza,''
I-00185, Roma, Italy \\
$^3$ Istituto dei Sistemi Complessi, ISC-CNR, Dep. Physics, Univ. Sapienza, P.le Aldo Moro 2, I-00185, Rome, Italy
}

\date{\today}

\begin{abstract}
We investigate Anderson localization of light as occurring in ultra-short excitations. 
A theory based on time dependent coupled-mode equations predicts universal features in the spectrum of the transmitted pulse. In particular, the process of strong localization of light is shown to 
correspond to the formation of peaks in both the amplitude and in
the group delay of the transmitted pulse. Parallel \emph{ab-initio} simulations made with Finite-Difference Time Domain codes and molecular dynamics confirm theoretical predictions,
while showing that there exist an optimal degree of disorder for the strong localization.
\end{abstract}

%\pacs{71.23.-k,42.25.Fx,82.70.Dd}

\maketitle
%%%%%%%%%%%%%%%%%%%%%%%%%%%%%%%%%%%%%%%%%%%%%%%%%%%%%%%%%%%%%%%%%%%%%%%%%%%%%%%%%%%%%%%%%%%%%%%
%%% INTRODUCTION
%%%%%%%%%%%%%%%%%%%%%%%%%%%%%%%%%%%%%%%%%%%%%%%%%%%%%%%%%%%%%%%%%%%%%%%%%%%%%%%%%%%%%%%%%%%%%
\emph{Introduction ---}
In the last years, progress in laser technology  has resulted in 
the generation of pulses with duration of few optical cycles \cite{Ranka, Bellini, Husakou, Dudley},
whose spectral content is an optical frequency comb, i.e.,
a discrete, regularly spaced, series of sharp lines in the region of optical frequencies. 
These new laser sources have
favoured many outcomes in fundamentals research and 
applications, ranging from optical frequency metrology \cite{Diddams00, Udem02, Diddams07} to pulse synthesis \cite{Shelton01,Savchenkov}.
In the same years, the systematic study of light propagation in disordered dielectrics 
has stirred particular attention 
\cite{Johnson03,Wiersma07,Conti08,Chabanov00,chabanov03,Wiersma97,Muskens08,Rockstuhl09}. 
In these systems the multiple scattering can become so strong that the diffusive 
approximation for the photon transport lapses and the randomness of the 
three-dimensional (3D) structure leads to the emergence of Anderson localizations,
i.e., long living exponentially localized electromagnetic resonances.
In several respects, the observation of these localizations 
is still largely debated \cite{Pinheiro04,Skipetrov06,Storzer06,Maret06,Swartz07,Conti07,Gentilini09}. 
One of the leading issue is determining universal features of light localization dynamics in order to infer
their excitation from the analysis of the spectrum (amplitude and phase) of the transmitted pulse.
In one-dimension (1D) transmission fluctuations can be related to exponentially localized or to delocalized necklace states,
both mediating a resonant like behaviour (see, e.g., \cite{Gil03,Bertolotti05}). In 3D, with reference to dielectric samples, emphasis has been given to the shape
of the tail of the transmitted pulse, that should display deviations from a diffusive exponential trend
in the presence of long living states \cite{Skipetrov06}.
However for a broad-band excitation such an analysis cannot be directly applicable because: 
(i) the diffusion approximation may not hold true in whole excitation spectrum, (ii) material dispersion should be accounted for.
In addition, when dealing with ultra-fast laser sources, one could argue on which is the Frequency Resolved
Optical Signal (FROG) in the presence of Anderson localization (see,e.g., \cite{Trebino08}), or which is the corresponding amplitude/phase profile in the spectrum and if they display universal distinctive features. In this respect, a time-resolved electromagnetic approach is 
unavoidable,  especially for dealing with a frequency dependent refractive index and broad-band excitations.\\
In this Manuscript, we investigate the response of a random medium to a frequency comb. 
We tackle this issue by Time Domain Coupled Mode Theory (TDCMT) \cite{HausBook}, and by an \emph{ab-initio} parallel numerical approach that employs
finite-difference time-domain (FDTD) codes combined with a 
Molecular Dynamics (MD) \cite{rapaport}, which allow to completely simulate the dynamics of Maxwell's equations (FDTD) inside realistic configurations of scatterers (MD).
We compare materials with different index contrast, and, for a strongly scattering medium, we vary the filling fraction as a measure of the degree of randomness.
Theory and simulations demonstrate that the transmitted pulse contains spectral universal features of the Anderson localization of light, which, according to the FDTD simulations, is mostly evident 
at a specific amount of disorder. This is the first ever reported first-principle study 
of the role of the 3D Anderson localization of light in the presence of a flat wide-band excitation,
and it is expected to stimulate further experimental and theoretical work.
Indeed, so far, most of the reported investigations dealt with a diffusive almost-monochromatic regime;
and a full Maxwell solution for a frequency-comb input was not reported before.
%%%FIGURE
\begin{figure}
\includegraphics[width=8.5cm]{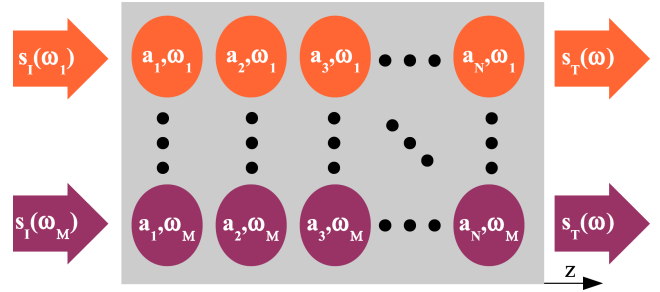}
\caption{(Color on line) Localization couplings and energy propagation scheme in the TDCMT.
\label{fig1}}
\end{figure}
%%%%%%%%%%

\begin{figure}
\includegraphics[width=8.5cm]{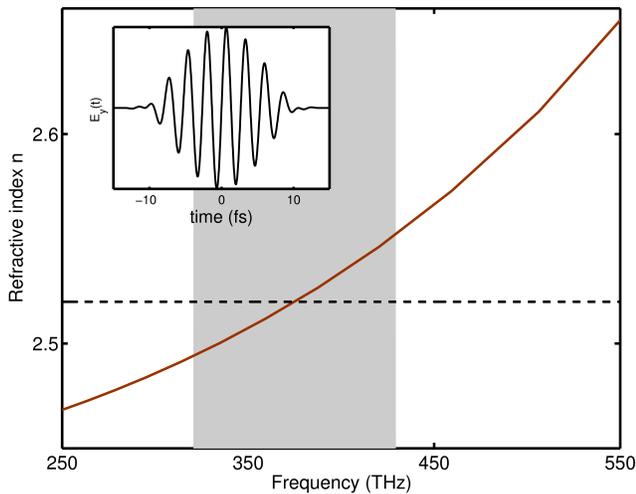}
\caption{(Color online). TiO$_2$ refractive index $n$ vs frequency,
as obtained from the Sellmeier equation after Devore \cite{Devore}.
The filled area corresponds to the frequency comb spectrum.
The inset shows the input pulse.
\label{fig1b}}\end{figure}

\emph{Theory ---}
In our theoretical picture the impinging laser pulse (propagation direction $z$, input surface at $z=0$) excites a series of Anderson localizations 
at different frequencies, which are expected in proximity of the surface of the sample;
energy is then transported in the sample by coupling between localized states at the same 
frequencies. We focus on the output signal at the input propagation direction, thus only those
localized modes that are overlapped with the beam contribute to the output, while the others behave as radiative losses.
The coupling between resonances at different frequencies will not contribute to the energy transport, indeed 
their beating will average out.
Such an approach can in principle applied to any dimensionality to describe transport of energy mediated by coupled 
localized resonances.
We consider $M$ chains of uncoupled oscillators, each corresponding
to a resonance $\omega_\nu$ ($\nu=1,2,...,M$) with amplitude
$a_{\nu,n}$ with $n=1...N$, where $N$ is the number of Anderson localizations at frequency $\omega_\nu$
overlapped with the input beam (Fig.\ref{fig1}) [we group resonances at the same position $z$ (within a localization length) into a single oscillator $a_{\nu,n}$]. The propagation of energy at frequency $\omega_\nu$ is then given by the linear TDCMT equations:
\begin{equation}
\label{andchain}
i\frac{d a_{\nu,n}}{d t}=\bigg(\omega_\nu-i\frac{1}{\tau_{\nu,n}}\bigg)a_{\nu,n}+\mu_{n-1,\nu}a_{n-1,\nu}+\mu_{n+1,\nu}a_{n+1,\nu},
\end{equation} 
with $\tau_{n,\nu}$ accounting both for the emission losses in directions different from the input beam and for
material absorption. The coeffcients $\mu$ in (\ref{andchain}) describe the coupling between modes and, in general, they
also account for material dispersion.
For $n=1$, the previous equations contain $\mu_{0,\nu}a_{0,\nu}\equiv\sqrt{2/\tau_e}s_I(\omega_\nu)$, which is the fraction of the 
signal coupled at frequency $\omega_\nu$, being $s_I(\omega_\nu)$ the amplitude of the supercontinuum at $\omega_\nu$ 
and $2/\tau_e$ the coupling loss rate
in $z$-direction (Fig. \ref{fig1}) \cite{HausBook}. Since the input beam will exponentially decrease 
with the sample length, we assume that only one single mode for each $\omega_\nu$ (the nearest to the input surface $z=0$) will be directly
excited by the input beam. For $n>1$, $a_{n-1,\nu}$ ($a_{n+1,\nu}$) is the Anderson localization at $\omega_\nu$ 
preceding (following) $a_{\nu,n}$. The transmitted signal is then $s_t=\mu_{N,\nu}a_{N,\nu}$, being $\mu$ the coupling coefficient with the output region. The transmission $T=s_t/s_i$ is then found by solving (\ref{andchain}) in the Fourier domain ($\frac{d}{dt}\rightarrow i\omega$):
\begin{align}
  \label{sol0}
\frac{ s_t(\omega)}{s_i(\omega_\nu)}=i^{N+2}\mu_{N,\nu}\prod_{k=1}^NG(k,\nu),
\end{align}
being $G(k,\nu)$ the following continued fraction:
\begin{align}
  \label{gexp}
  G(k,\nu)=\begin{cases}
    \frac{\mu_{k-1,\nu}}{i(\omega-\omega_\nu)+\frac{1}{\tau_k}+\mu_{k+1,\nu} G(k+1)}, &k\le N\\
    0, &k>N
\end{cases}.
\end{align}
Universal properties of the system can be demonstrated from (\ref{sol0})-(\ref{gexp}),  by looking at the two physical observables: the transmittance $\lvert T\rvert^2$ and the group delay $\frac{d\phi}{d\omega}$ (with $\phi$ being the phase of $T$). We begin by demonstrating with mathematical induction the property $\Im[G(m,\nu)]\lvert_{\omega=\omega_\nu}=0$. This is trivially fulfilled for $G(N)$, and also for $G(k)$:
\begin{align}
  \Im[G(k)]=\frac{-\mu_{k-1,\nu}\big[(\omega-\omega_\nu)+\mu_{k+1,\nu}\Im[G(k+1)]\big]}{\lvert i(\omega-\omega_\nu)+\frac{1}{\tau_k}+\mu_{k+1,\nu}G(k+1)\rvert^2},
\end{align}
if $\Im[G(k+1,\nu)]\lvert_{\omega=\omega_\nu}=0$. In the same way we can verify that $\Re[\frac{dG(k,\nu)}{d\omega}]\lvert_{\omega=\omega_\nu}=0$ and, as a corollary, $\Im[G(k,\nu)^2]\lvert_{\omega=\omega_\nu}=0$, $\Re[G(k,\nu)^2\frac{dG(k,\nu)}{d\omega}]\lvert_{\omega=\omega_\nu}=0$ and $\Im[\frac{d^2G(k,\nu)}{d\omega^2}]\lvert_{\omega=\omega_\nu}=0$. By induction we then obtain:
\begin{align}
\label{res}
  &\frac{d\lvert G(k,\nu)\rvert^2}{d\omega}\lvert_{\omega=\omega_\nu}=0,
  &\frac{d^2\phi_{k,\nu}}{d\omega^2}\lvert_{\omega=\omega_\nu}=0,
\end{align}
being $\phi_{k,\nu}$ the phase of the continued fraction $G(k,\nu)$. Correspondingly $|T|^2\propto\prod_k\lvert G(k,\nu)\rvert^2$  and $\phi\propto\sum_k\phi_{k,\nu}$; hence, owing to Eq.(\ref{res}), we are able to say that in the presence of Anderson localization occurring with multiple resonances, as produced, e.g., by broadband excitations, for each resonance $\omega_\nu$ the transmission spectrum exhibit extrema in the amplitude and the group delay.
\\\emph{Sample and numerical setup---}
The previous general arguments have a universal character and applies to a variety of different situations; in order to make a specific example and verify
the validity of our theoretical approach we resorted to first-principle FDTD parallel simulations. 
To consider a realistic sample of disordered colloidal system we used
 MD simulations, which furnish a distribution of 8000  mono-dispersed spheres with radius $r$ and filling fraction $\Phi$ in air interacting with a Lennard-Jones potential.
We first consider two different materials with $r\sim150$~nm at a filling fraction $\Phi\sim 0.55$: Silica SiO$_2$ glass ($n\cong1.5$) and 
Titanium Dioxide Ti$O_2$ in its Rutile form ($n\cong2.5$). With reference to realistic frequency comb generators, 
we tailored the temporal profile of the impinging pulse so that 
its spectral content is centred around the 
carrier frequency $f_0=375$~THz ($\lambda_0=800$~nm),
and regularly spaced between $320$~THz and $430$~THz 
by a repetition rate $f_{rep}\sim90$~GHz. 
The input pulse $E(t)=\sum_n \sin (2\pi f_nt)$,
where the sum is performed over $1000$ equi-spaced frequencies $f_n$ (Fig.\ref{fig1b}).

In our numerical experiments we take a train of ultra-short, y-polarized  pulse impinging on 
the x-y face of the sample at normal incidence. The input spatial profile
is Gaussian TEM$_{00}$  with waist $w_0=1$~$\mu$m. 
$10$~ps runs take one hour by using $1024$ processors on an IBM system.
%%%FIGURE
%%%%%%%%%%

For each simulations, we collect: i) the spatial profiles of the electric field at the output plane
of the sample; ii) the electromagnetic energy $\mathcal{E}$ and 
iii) the corresponding spectrum. Afterwards we calculate: a) the group delay as 
the derivative, with respect to frequency, of the phase of Fourier transform of 
the electric field; b) the frequency-resolved optical gate (FROG) signal 
$I_{FROG}(\omega,\tau)$ as:
\begin{equation}
I_{FROG}(\omega,\tau)\propto\vert\int_{-\infty}^{+\infty}E_y(t)E_y(t-\tau) e^{-i\omega t} dt\vert^2
\label{eq.2}\end{equation}where $E_y(t)$ is the output signal, which corresponds to the complex signal $s_t$ from
TDCMT above.
%%%%%FIGURE
\begin{figure}
\includegraphics[width=8.3cm]{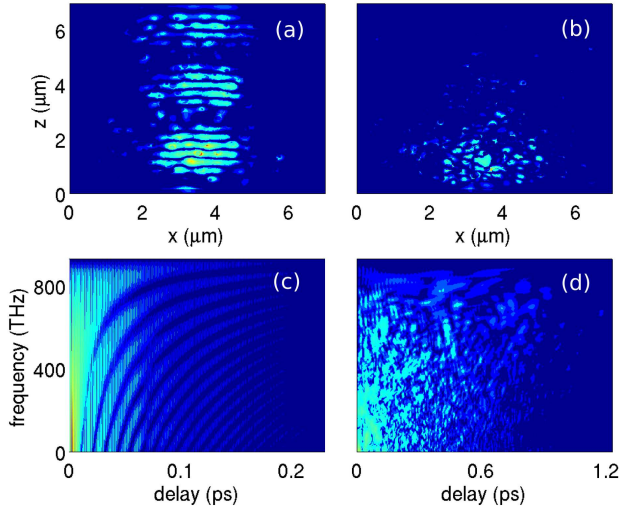}
\caption{(Color on line) Slices $(x,z)$ (in the middle of y axis) of the electromagnetic energy 
$\mathcal{E}$, resulting from dispersive simulations  of SiO$_2$ (a) and 
$TiO_2$ (b).\label{fig3}
Corresponding FROG signal of the transmitted pulse ($E_y$) 
for SiO$_2$ (c) and for $TiO_2$ (d).}
\end{figure}

\emph{Numerical results ---}
The first signature of the existence of localized modes 
in TiO$_2$ is found by looking at Fig.\ref{fig3}a,b , where 
we notice a drastic change in the spatial profile of the Energy $\mathcal{E}$ from extended states in SiO$_2$ (\ref{fig3}a) to localized states in 
TiO$_2$(\ref{fig3}b).
To characterize the presence of localized modes, we compute the FROG signal $I_{FROG}(\omega,\tau)$ \cite{Trebino93,Trebino94,Trebino08}.
Figure \ref{fig3} shows $I_{FROG}(\omega,\tau)$ calculated 
by the Eq.(\ref{eq.2}): panel (c) corresponds to the field $E_y(t)$ coming out by SiO$_2$, while panel (d) refers to TiO$_2$. 
The FROG signal is able to detect if the strong localization is set in the medium. Indeed, 
the pattern in the right panel (d) displays a spectrogram in which all the frequencies
(reported in the y-axis) survive for longer time ($\sim1.2$~ps) with respect to the 
case of Silica glass in panel (c) ($\sim0.2$~ps) [Note that the FROG spectrum is doubled with respect to the field spectrum]. This observation is in agreement
with the emergence for TiO$_2$ of several localized modes with a long lifetime,
corresponding to amplitude peaks and longer delays.
Comparison with our theoretical predictions can
be tackled through the spectral analysis on
the Fourier transform $T(\omega)$ of the output pulse $E_y(t)$:
\begin{equation}
T(\omega)=A(\omega)\exp[i\frac{\omega n(\omega)}{c}L]=A(\omega)e^{i\psi(\omega)}\text{.}
\label{transmission}
\end{equation}
The derivative $d\psi/d\omega$ is equal to the product between 
the inverse of group velocity $v_g=d\omega/dk$ and the distance $L$ covered 
by the incident pulse at the end of the sample, that is the group delay $\tau_g=L/v_g$. 
%%%%%FIGURE
\begin{figure}  
\includegraphics[width=8.5cm]{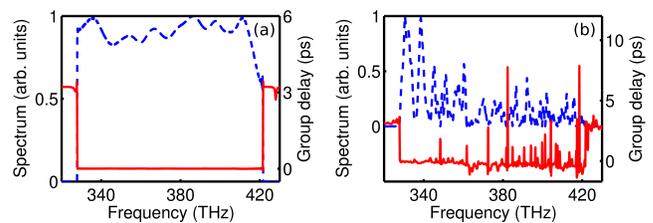}
\caption{(Color on line) Squared modulus of the Fourier transform of $E_y$ (dashed line, left axis) 
and group delay (continuous line, right axis) for SiO$_2$ (a) and $TiO_2$ (b).
The vertical axis of the group delay is shifted such that
the zero corresponds to the exit time of the low frequency side of the input spectrum.
\label{fig5}}
\end{figure}
%%%%%END FIGURE

Figure \ref{fig5} shows the squared modulus $A^2(\omega)=|T(\omega)|^2$ 
and the group delay $\tau_g(\omega)$: we compare the results for SiO$_2$  (Fig. \ref{fig5}a) with those TiO$_2$ (Fig.\ref{fig5}b).
The transmission gives for the transport mean free path $l\cong0.6$~$\mu$m for TiO$_2$ ($kl\cong5$) and $l=10$~$\mu$m for SiO$_2$ ($kl\cong80$).
Localization is accompanied in TiO$_2$ by the formation of several peaks for $A^2(\omega)$ and $\tau_g(\omega)$
(absent for SiO$_2$) in agreement with theoretical predictions. Note that due to the multiple couplings (i.e., from a resonant state
to another or mediated by back-coupling to the input beam), both minima and maxima for the delay and the
amplitude are obtained (corresponding to mixed side-coupling and forward-coupling regimes for optical cavities \cite{HausBook}).
Such a response is a universal feature that we found for any
considered disorder realization. We also compared dispersive and non-dispersive materials, which resulted
into quantitatively different distribution of peaks (not shown), while retaining for both cases the predicted features.

The link between the amount of disorder and the onset of Anderson localization is a key issue. Previous
numerical investigations for 3D photonic crystals \cite{Conti07} have shown that the localization length is minimal for a specific degree of disorder.
Notwithstanding the fact that we are considering a completely disordered regime, well beyond the perturbative regime considered in \cite{Conti07}, 
here one expects that such an optimal condition corresponds to the largest fluctuations of the group delay.
Hence we compared various TiO$_2$ structures exhibiting different filling fractions, as obtained by varying
the radius of the (eventually overlapping, as in porous media, see insets in figure \ref{fig4} below) spheres. The filling fraction $\Phi$ is adopted here as a measure
of the degree of disorder, indeed, for low (high) $\Phi$ the structure is mainly by air (TiO$_2$),
and can be taken as an ordered, homogeneous-like, system: an optimal value for $\Phi$ must exist for Anderson mediated
electromagnetic resonances.
\\In figure \ref{fig4} we show the spectra attained for three difference values of $\Phi$,
the group delay are clearly more pronounced in the panel (b). 
In the bottom panel we show the maximum group delay versus the filling fraction,
that display the greatest values in the range $\Phi\in[0.3,0.5]$,
corresponding to optimal disorder for the localization.
Note that a modulated amplitude is also present for large $\Phi$ (right panel),
but the attained delays are much smaller than for $\Phi\cong0.4$ (or vanishing as in figure \ref{fig5}a above);
this is related to boundary reflections (for large $\Phi$ the whole structure is a cube resonator) and ``sub-critical'' (i.e. weakly) localized stated.
Indeed, Anderson localization is signaled by the presence of large modulations in {\it both} the transmission and in the delay; at variance with the 1D case 
(see,e.g., \cite{Gil03}), in 3D a modulated trasmission does not necessarily correspond to an Anderson localization.
On the other hand, the narrow and sharp flat region for the delay versus $\Phi$ in Fig.\ref{fig4} signals the existence of two phases (i.e., diffusive 
and localized); 
large delays for a specific disorder are directly related to the onset of the strong-localization, which also results in the 
FROG signal in Fig.\ref{fig3}. We stress that the attained maximum delays for the various $\Phi$ in the ``optimal region'' correspond
to different resonances, thus denoting a largely frequency dependent regime, if compared to a shallow diffusive propagation.
%%%%%FIGURE4
\begin{figure}
\includegraphics[width=8.5cm]{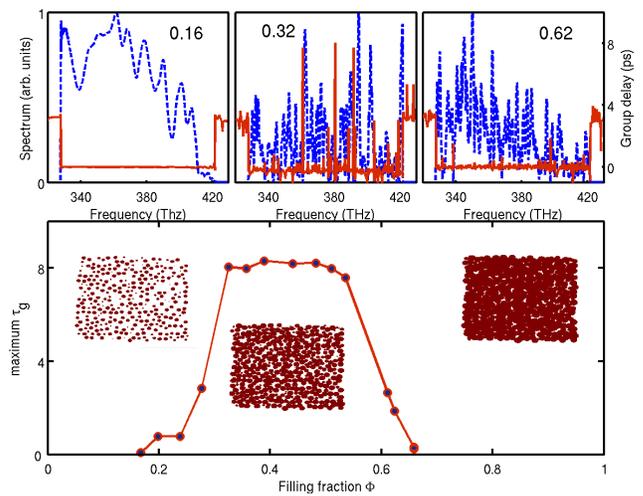}
\caption{(Color on line). 
(Top panels) Spectral response (squared modulus,left scale, dashed line, and 
group delay, right scale, continuous line) for different filling fractions for TiO$_2$.
(Bottom) Maximum attained group delay versus filling fraction;
the insets shows the material distribution, corresponding to the top panels.
\label{fig4}}
\end{figure}
%%%%%%%%%%%

\emph{Conclusions ---}
We have reported on the universal features of the Anderson localization of light 
in the presence of an ultra-wide band excitation.
We have shown that strong localization phenomena can be discriminated by the appearance of a resonant like
response in the amplitude and by the group delay versus frequency,
which is maximum at a specific disorder that can be interpreted as the optimal configuration for
trapping light in a random material.

\emph{Acknowledgements ---}
We acknowledge support from the INFM-CINECA initiative 
for parallel computing and CASPUR.
The research leading to these results has received
funding from the European Research Council under the European Community’s
Seventh Framework Program (FP7/2007-2013)/ERC grant agreement n.201766. Andrea Fratalocchi's 
research is supported by Award No. KUK-F1-024-21 (2009/2012), 
made by King Abdullah University of Science and Technology (KAUST).

%\bibliography{comb}

\end{document}